\documentclass{article}

\usepackage{epsfig}
\usepackage{subfigure}
\usepackage{calc}

\usepackage{amsfonts}
\usepackage{amsmath}
\usepackage{amssymb}
\usepackage{latexsym}

\begin{document}
\title{A Simple Attack on Some Clock-Controlled Generators}
\date{}
\author{P. Caballero-Gil$^{(1)}$, A. F\'uster-Sabater$^{(2)}$\\
{\small (1) Department of Statistics, Operations Research and Computing,}\\
{\small Faculty of Mathematics, University of La Laguna, 38271 Tenerife, Spain.}\\
{\small  pcaballe@ull.es}\\
{\small (2) Institute of Applied Physics (C.S.I.C.), Serrano 144, 28006 Madrid, Spain. }\\
{\small amparo@iec.csic.es}}
\maketitle
\begin{abstract}
We present a new approach to edit distance attacks on certain
clock-controlled generators, which applies basic concepts of Graph
Theory to simplify the search trees of the original attacks in
such a way that only the most promising branches are analyzed. In
particular, the proposed improvement is based on cut sets defined
on some graphs so that certain shortest paths provide the edit
distances. The strongest aspects of the proposal are that the
obtained results from the attack are absolutely deterministic, and
that many inconsistent initial states of the target registers are
recognized beforehand and avoided during search.

\end{abstract}
\section{Introduction}
\footnotetext{Work developed in the frame of projects:
HESPERIA supported by Centro para el Desarrollo Tecnol\'{o}gico
Industrial under program CENIT, and TIN2008-02236/TSI supported by the Spanish Ministry of Science and Innovation and FEDER Funds.\\
Computers \& Mathematics with Applications. Vol. 58 ,  Is. 1  (July 2009) pp. 179-188\\
DOI: 10.1016/j.camwa.2009.03.103}

The primary goal in the design of stream ciphers is to generate
long pseudorandom keystream sequences from a short key in such a
way that it is not possible to reconstruct the short key from the
keystream sequence. This work focuses on stream ciphers based on
Linear Feedback Shift Registers ($LFSRs$), and more particularly
on Shrinking \cite{CKM94} and Alternating Step \cite{GM04}
generators. Both generators produce keystream sequences with high
linear complexity, long period and good statistical properties
\cite{GC89}, and have been thoroughly analyzed in several papers
such as \cite{SGD98} and \cite{K03}.

Most types of cryptanalysis on stream ciphers are performed under
a known plaintext hypothesis, that is to say, it is assumed that
the attacker has direct access to the keystream output from the
generator \cite{J98}. The computational complexity of such attacks
is always compared with the complexity of the exhaustive search,
and if the former is smaller, then the cipher is said to be
broken. Although this theoretical definition can look useless, in
fact it is very important for the development and understanding of
the security of stream ciphers because many times it reveals
weaknesses that might lead to practical attacks.

The main idea behind this paper is to propose an optimisation approach that leads to a deterministic and exponential improvement of the complexity
of a known plaintext divide-and-conquer attack
consisting of three steps:

1.- Guess the initial state of an $LFSR$ component of the
generator.

2.- Try to determine the other variables of the cipher based on
the intercepted keystream.

3.- Check that guess was consistent with  observed keystream
sequence.

This three-step attack was first proposed in \cite{GM91} and
\cite{GP93} by means of a theoretical model and a distance
function known as Levenshtein or edit distance. Furthermore,
\cite{JG02} proved that when the length of the intercepted
sequence is large enough, the number of candidate initial states
in such an attack is small.

The approach considered in this work may be seen as an extension
of the constrained edit distance attack to clock-controlled
$LFSR$-based generators presented in \cite{PF04} and generalized
in \cite{CF05}. Our main aim here is to investigate whether the
number of initial states to be analyzed can be reduced. This
feature was pointed out in \cite{G98} as one of the most
interesting problems in the cryptanalysis of stream ciphers.
According to the original method, the attacker needs to traverse
an entire search tree including all the possible $LFSR$ initial
states. However, in this work the original attack is improved by
simplifying the search tree in such a way that only the most
efficient branches are retained. In order to achieve such a goal,
cut sets are defined in certain graphs that are here used to model
the original attack. This new approach produces a significant
improvement in the computing time of the original edit distance
attack since it implies a dramatic reduction in the number of
initial states that need to be evaluated. This quantitative
improvement is well established through implementation for the
specific cases of Shrinking and Alternating Step Generators.
Furthermore, it is remarkable that, unlike previous attacks, the
results obtained with the proposal of his work are fully
deterministic.

This work is organized as follows. Section 2 introduces basic
definitions of Shrinking and Alternating Step Generators and
essential concepts regarding edit distances. In Section 3, some
ideas for an efficient initial state selection method are given
that allow describing a method for deducing a threshold value for
the computation of the edit distance. Section 4 presents the full
description of the proposed attack. Finally, Section 5 contains
specific details for the cases of Shrinking and Alternating
Generators and Section 6 provides several conclusions.

\section{Preliminaries}

The Shrinking Generator ($SG$) is a nonlinear combinator based on
two $LFSRs$, introduced in 1993 by Coppersmith, Krawczyk and
Mansour \cite{CKM94}. In this generator the bits produced by one
$LFSR$, denoted by $S$, are used to determine whether the
corresponding bits generated by the second $LFSR$, denoted by $A$,
are used as part of the overall keystream or not.

The Alternating Step Generator ($ASG$) is a nonlinear combinator
based on three $LFSRs$, proposed in 1987 by G$\ddot{u}$nther
\cite{G88}. According to this generator each bit produced by one
of the three $LFSRs$, denoted by $S$, is used to determine the
output from the other two $LFSRs$ $A$ and $B$. In this work we use
a modified and equivalent version of the original $ASG$ defined as
follows. If $S$ produces a 1, then the output bit of the generator
is the output bit from the $LFSR$ $A$, which is clocked.
Otherwise, if $S$ produces a 0, then the output bit of the
generator is the output bit from the $LFSR$ $B$, which is clocked.

The notation used within this work is as follows. The lengths of
the $LFSRs$ $S$, $A$ and $B$ are denoted respectively by $L_S$,
$L_A$ and $L_B$. Their characteristic polynomials are respectively
$P_S(x)$,  $P_A(x)$ and $P_B(x)$, and the sequences they produce
are denoted by  $\{s_i\}$, $\{a_i\}$ and $\{b_i\}$. The output
keystream is $\{z_j\}$.

Despite their simplicity and the large number of  published
attacks \cite{DV05} \cite{G05} \cite{GKW06} \cite{ZWFB05}
\cite{Jab96}, both generators remain remarkably resistant to
practical cryptanalysis because there are no known attacks that
may be considered efficient enough when the LFSRs are too long for
exhaustive search. Each reference above may be described either as
a theoretical attack because it requires hard hypothesis or/and
the obtained results are probabilistic, or as an attack launched
against the hardware implementation of the generator. One of the main advantages of this work is that it proposes a new deterministic approach to the cryptanalysis of
$LFSR$-based stream ciphers.

The edit or Levenshtein distance is the minimum number of
elementary operations (insertions, deletions and substitutions)
required to transform one sequence $X$ of length $N$ into another
sequence $Y$ of length $M$, where $M\leq N$. Some applications of
the edit distance are file checking, spell correction, plagiarism
detection, molecular biology and speech recognition \cite{Gus97}. The dynamic
programming approach (like the shortest-distance graph search and
Viterbi algorithm) is a classical solution for computing the edit
distance matrix where the distances between prefixes of the
sequences are successively evaluated until the final result is
achieved. When applying an edit distance attack on a
clock-controlled stream cipher, the objective is to compute the
initial state of a target $LFSR$ that is a component of the
attacked generator. As in Viterbi search, this problem has the
property that the shortest path to a state is always part of any
solution of which such a state is part. We will be able to see
this fact quite clearly by the formalization of the algorithm as a
search through a graph.

Clock-controlled registers are said to work with constrained
clocking when a restriction exists on a maximum number of times
that the register may be clocked before an output bit is produced.
For these registers, attacks based on a so-called constrained edit
distance have been proposed and analyzed in \cite{GM91} and
\cite{GP93}.

In this work, two different possible models for the attacked
generator, shown respectively in Fig. 1 and Fig. 2, are
considered. In both cases it is assumed that the feedback
polynomial of the target $LFSR$ is known. According to the first
model, it is assumed that $Y=\{y_n \}$ is an intercepted keystream
segment of length $M$, which is seen as a noisy decimated version
of a segment $X=\{x_n \}$ of length $N$ produced by a target
$LFSR$. On the other hand, according to the second model, it is
assumed that $X=\{x_n \}$ is an intercepted keystream segment of
length $N$, which is seen as a noisy widened version of a segment
$Y=\{y_n \}$ of length $M$ produced by a target $LFSR$. In this
latter case, insertions in the sequence $Y$ are indicated by two
sequences $S$ and $B$ so that $S$ points the locations where the
bits of $B$ must be inserted. The simplest examples represented by
these theoretical models are $SG$ and $ASG$, respectively.
Furthermore, in both cases it is not necessary to consider noise
to model the generator.

\begin{figure}[ht]
  \begin{center}
     \includegraphics[bb=0 0 1349 391,width=4in]{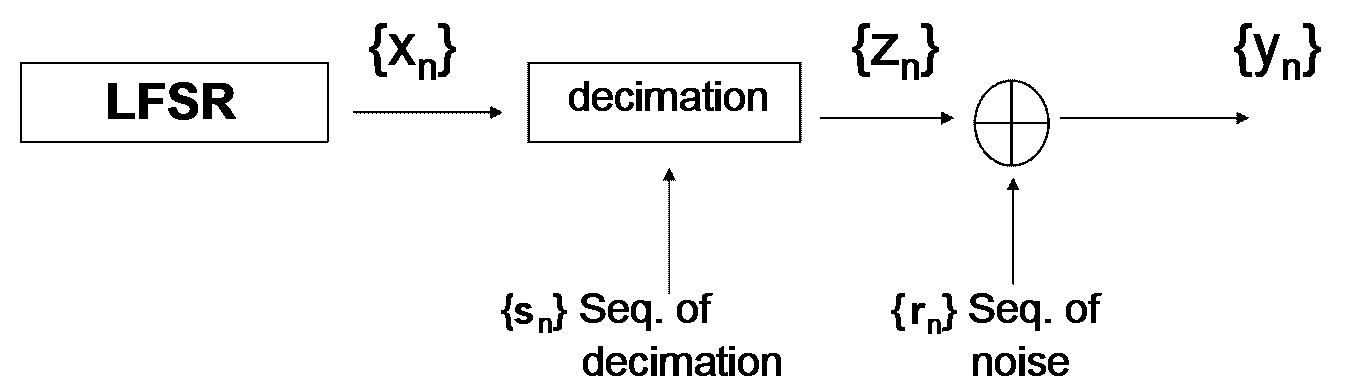}
      \end{center}
  \caption{Theoretical model with decimations}
\end{figure}

\begin{figure}[ht]
  \begin{center}
     \includegraphics[bb=0 0 1349 391,width=4in]{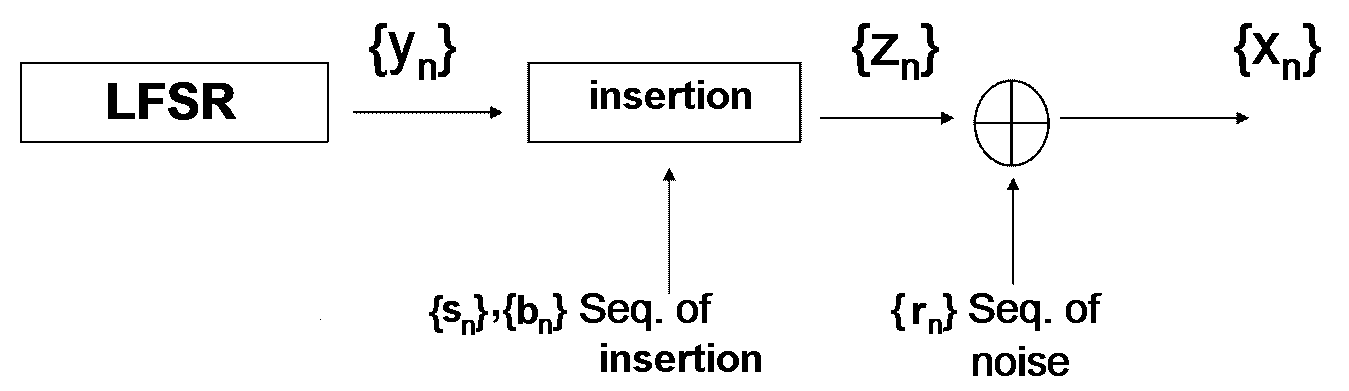}
      \end{center}
  \caption{Theoretical model with insertions}
\end{figure}

The main objective of the attack according to these models will be
to deduce some initial state of the target $LFSR$ that allows
producing an intercepted keystream sequence through decimation or
insertion, respectively, without knowing the decimation or
insertion sequences. In either case, the attack is considered
successful if only a few initial states are identified.

The use of these models implies that the known plaintext attack is
applicable not only to those generators that fit exactly such a
model but also to other sequences produced by more complex
generators that generalize the given description. In this latter
case, the attack would provide a simpler equivalent description of
the original attacked generator.

An essential step in edit distance attacks is the computation of
edit distance matrices $W=(w_{i,j}), i=0,1,\ldots,N-M,
j=1,2,\ldots,M $ associated each one with a couple of sequences
$X$ and $Y$ where $Y$ is the intercepted keystream sequence and
$X$ is a $LFSR$ sequence produced by one possible initial state.

It is remarkable that the computation of the edit distance matrix
requires choosing between both models the one that better fits the
attacked generator. If it is the first model, the intercepted
sequence is $Y$ while $X$ is the candidate sequence. Otherwise, if
it is the second model, then the intercepted sequence is $X$ while
$Y$ is the candidate sequence. Also note that from the computation
of the edit distance between $X$ and $Y$, the edit sequences that
are computed in the first case correspond to decimation sequences
while in the second case they correspond to insertion sequences.

Some of the parameters of such a matrix are described below.
Firstly, its dimension is $(N-M+1)\cdot M$. Furthermore, its last
column gives the edit distance between $X$ and $Y$ thanks to the
value $\min_{i=0,\ldots,N-M}\{w_{i,M}+N-M-i\}$. Lastly,  each
element of the matrix but the last column $w_{i,j},\
i=0,\ldots,N-M,\  j=1,\ldots, M-1$   corresponds exactly to the
edit distance between prefix sub-sequences $x_1,x_2
,\ldots,x_{i+j}$ and $y_1,y_2,\ldots,y_j$. The edit distance
between prefix sub-sequences $x_1,x_2,\ldots,x_{i+M}$ and $Y$ are
given by $ w_{i,M}+N-M-i,\  i=0,\ldots, N-M$.

In the edit distance attack here analyzed only deletions and
substitutions are allowed. Consequently, each element $w_{i,j}$ of
the edit distance matrix $W$ may be recursively computed from the
elements of the previous columns according to the formulas in
Equation (1), which depend exclusively on the coincidence or
difference between the two bits $x_{i+j}$ and $y_j$.

\begin{center}
$w_{i,1}= P_i(x_{i+1},y_1),\ i=0,\ldots,N-M$

$w_{0,j}= w_{0,j-1}+P_0(x_j,y_j),\ j=2,\ldots,M$

$w_{i,j}=\min_{k=0,\ldots, i} \{w_{i-k,j-1} +P_k(x_{i+j}, y_j)\},\
i=1,\ldots,N-M,\ j=2,\ldots,M$

$P_k(x_{i+j},y_j)=\left\{ \begin{array} {ccc}
  k &
 if & x_{i+j}=y_j\\
  k+1 & if
& x_{i+j}\neq y_j

\end{array} \right ., k=0,...,i$
\begin{equation}\end{equation}
\end{center}

$P_k(x_{i+j},y_j)$ gives the cost of the deletion of $k$ bits
previous to $x_{i+j}$ plus its substitution by its complementary
if $x_{i+j}\neq y_{j}$. Note that a maximum length $k$ of possible
runs of decimations is assumed for constrained edit distance
matrices. It is also remarkable that at each stage the minimum has
to be obtained in order to extend the search at a next stage,
which implies the need to maintain a record of the search in the
same way that Viterbi algorithm saves a back pointer to the
previous state on the maximum probability path.

In order to avoid the computation of the edit distances for all
possible initial sequences; in the following we propose a
graph-theoretic approach so that the computation of edit distances
may be seen as a search through a basic graph. Such a basic graph
is a directed rooted tree where each non-root vertex $(i+j,j),
i=0,1,…,N-M; j=1,2,…,M$, indicates a correspondence between the
bits $x_{i+j}$ and $y_j$ and each edge indicates either a deletion
of the bit $x_{i+j}$ when $j=0$, or a possible transition due to a
deletion (D) or a substitution (S), in the remaining cases. In
this way, the computation of edit distances consists in finding
the shortest paths through the graph in Fig. 3.


\begin{figure}[htb]
  \centering
  \includegraphics[bb=0 0 945 418,width=4in]{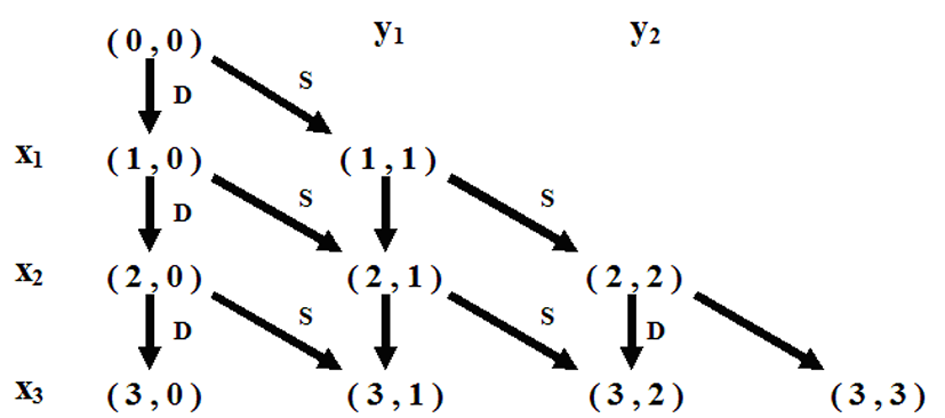} 
  \caption{Group Detection}
  \label{Fig:Detection}
\end{figure}

For the description of our improvement, we now define a new
weighted directed graph, here called induced graph, where the
costs of shortest paths come directly from the elements of the
matrix $W$. This induced graph is computed from the basic graph
shown in Fig. 3 as follows. If we eliminate vertical edges in the
graph of Fig. 3 by computing the partial transitive closure of
every pair of edges of the form $((i+j-2, j-1),(i+j-1,j-1))$ and
$((i+j-1,j-1),(i+j,j))$ and substituting them by the edge
$((i+j-2, j-1),(i+j,j))$, we get the graph shown in Fig. 4, which
is here called induced graph.

\begin{figure}[ht]
  \begin{center}
     \includegraphics[bb=0 0 945 444,width=4in]{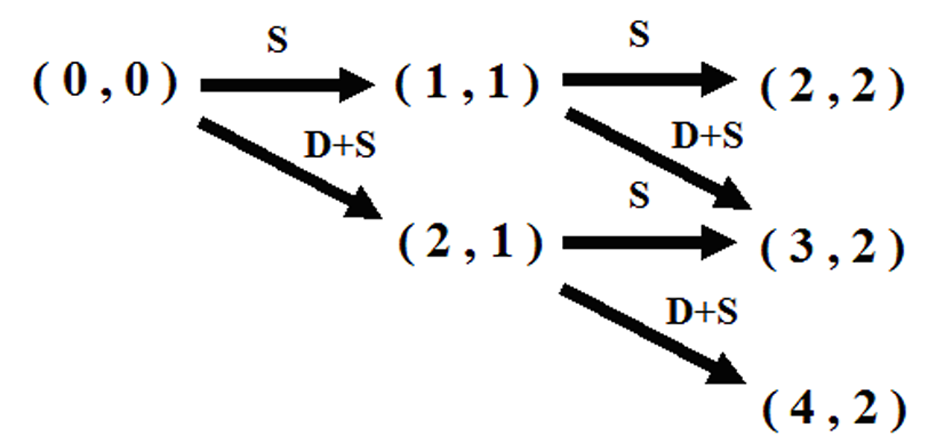}
      \end{center}
  \caption{Induced graph}
\end{figure}

In this graph there are as many vertices as elements in the matrix
$W$, plus an additional source and an additional sink. On the
other hand, the directed edges in this induced graph are defined
from the computation of the edit distances described in Equation
(1), plus additional edges joining the source with the vertices
associated to the first column of $W$ and additional edges joining
the vertices associated to the last column of $W$ with the sink.
For instance, the induced graph corresponding to a constrained
edit distance matrix with runs of decimations of maximum length 1
has $(N-M+1)\cdot (2M-N+2)$ vertices and $2 \cdot (N-M+1) \cdot
(2M-N+2)-M-3$ edges. Moreover, edges in the induced graph have
different costs depending on the specific pair of sequences $X$
and $Y$, and particularly on the coincidences between the
corresponding bits of both sequences, as described in Equation
(1). Note that in the induced graph, the shortest paths between
the source and the sink give us the solution of the cryptanalytic
attack through the specification of both decimation and noise
sequences that can be extracted from them.

$Example:$ For an intercepted keystream sequence $Y$:1101011 of
length $M$=7 and a candidate sequence $X$:1110110111 of length
$N$=10, the constrained edit distance matrix with runs of
decimations of maximum length 1 is:

$W= \left(
\begin{array} {ccccccc}
  0 & 0 & 1 & 2 & 3 & - & - \\
   1 & 1 & 1 & 1 & 2 & 3 & - \\
   - & 3 & 3 & 2 & 2 & 2 & 2 \\
   - & - & 5 & 5 & 4 & 3 & 3
\end{array}\right)$.

The graph induced by this matrix is shown in Fig. 5 where there
are exactly 24 vertices and 38 edges of which the ones belonging
to the 18 shortest paths between the source and the sink are
remarked in grey.

\begin{figure}[ht]
  \begin{center}
     \includegraphics[bb=0 0 945 491,width=4in]{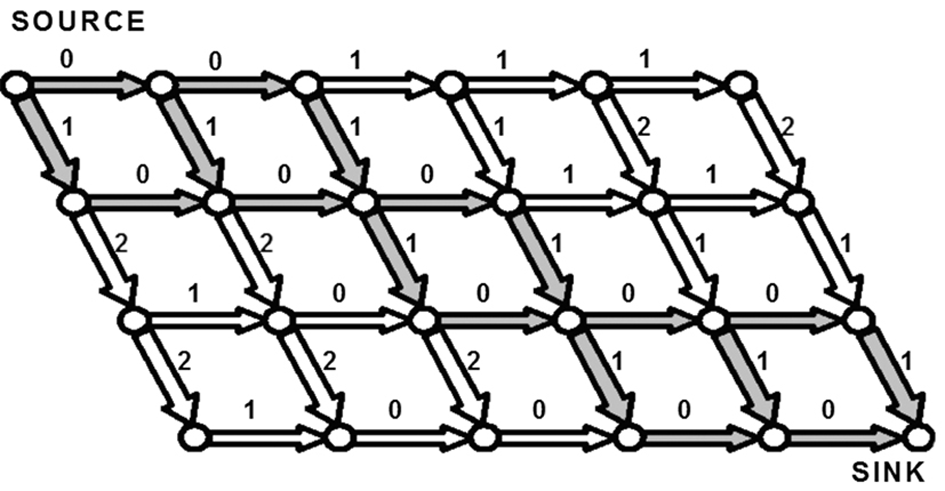}
      \end{center}
  \caption{Shortest paths}
\end{figure}

For each one of those 18 optimal paths, we obtain a possible
solution to the cryptanalysis. The computation of 18 decimation
sequences $S$ may be deduced from the above graphical
representation in the following way. A horizontal edge in an
optimal path is interpreted as a 0 in a deduced decimation
sequence (that is to say, no deletion of the corresponding bit)
whereas each oblique edge in the path gives a 1 as decimation bit
(corresponding to the deletion of the corresponding bit).

$S=\{s_n\}: \left\{ \begin{array}{cc}
 0111011011: & Solution 1 \\
 0111011101: & Solution 2 \\
 0111011110: & Solution 3 \\
 0111101011: & Solution 4\\
 0111101101: & Solution 5 \\
 0111101110: & Solution 6 \\
 1011011011: & Solution 7 \\
 1011011101: & Solution 8 \\
 1011011110: & Solution 9 \\
 1011101011: & Solution 10 \\
 1011101101: & Solution 11 \\
 1011101110: & Solution 12 \\
 1101011011: & Solution 13 \\
 1101011101: & Solution 14 \\
 1101011110: & Solution 15 \\
 1101101011: & Solution 16 \\
 1101101101: & Solution 17 \\
 1101101110: & Solution 18
 \end{array} \right . $

\section{ Search of Promising Initial States }

The main idea behind the method shown in this Section comes
directly from the association between bits $x_{i+j}$ and edges of
the induced graph. Since the calculation of the minimum edit
distance implies the computation of some shortest path in such a
graph, cut sets between the source and the sink in the induced
graph may be useful in order to define a set of conditions for
candidate sequences so that it is possible to establish a minimum
threshold edit distance. In this way, once an intercepted sequence
fulfills some of those stated conditions, the cost of the
corresponding cut set can be guaranteed to be minimal for some
possible candidate sequence, what has direct consequences on the
costs of the shortest paths, that is to say, on the edit
distances.

In this way, as soon as an intercepted sequence fulfills some
specific condition defined below, and this fact allows the
description of a candidate and feasible initial sequence, we will
know that such an initial sequence will provide us with a useful
upper threshold for the edit distance and even in many cases, such
a sequence will be a minimum edit distance sequence.

The specific cut sets that we have used for the numerical results
shown in this work are defined as follows. Each cut set $C_{i+j},
2\leq i+j\leq N-1$ contains:
\begin{enumerate}
    \item The set of all the arcs corresponding to the vertex $x_{i+j}$.
    \item All those edges corresponding to bits $x_{w}$ with $ w>i+j$ whose output vertex is one of the output vertices of the former set.
\end{enumerate}

For the first model, these cut sets may be characterized by
several independent conditions on the intercepted sequence $Y$
that may be used to guarantee a decrease on the edit distances of
different candidate sequences $X$. After having checked each
hypothesis separately, the tools used to check both sets of
conditions on candidate sequences $X$ are described in terms of a
pattern that is made out of independent bits of $X$ according to
the formulas in Equation (2).

\begin{description}
\item If $\forall j:2, 3,\ldots, N-M+1;\ y_1=y_2=
\cdots=y_{j}$ then $y_j=x_1=x_2= \cdots=x_{j+N-M}$
\item If $\forall j:N-M+2, N-M+3,\ldots, M; y_{M-N+j}=
\cdots=y_{j-1}=y_j$ then $\ y_j=x_{j}=x_{j+1}= \cdots=x_{j+N-M}$
\item If $ \forall j: M+1, M+2,\ldots, N-1;\  y_{M-N+j}=
\cdots =y_{M-1}=y_M$ then $y_M= x_{j}=x_{j+1}=\cdots=x_{N}$
\end{description}\begin{equation}\end{equation}

For the second model, the cut sets may be characterized by
different independent conditions on the intercepted sequence $X$
that may be used to guarantee a decrease on the edit distances of
candidate sequences $Y$. After having checked each hypothesis
separately, the tools used to check both sets of conditions on
candidate sequences $Y$ are described in terms of a pattern that
is made out of independent bits of $Y$ according to the formulas
in Equation (3).

\begin{description}

\item If $\forall j:2, 3,\ldots, N-M+1;\ x_1=x_2= \cdots=x_{j+N-M}$ then $x_1=y_1=y_2=\cdots=y_{j}$

\item If $\forall j:N-M+2, N-M+3,\ldots, M; x_{j}=x_{j+1}=
\cdots=x_{j+N-M} $ then $\ x_j= y_{M-N+j}= \cdots=y_{j-1}=y_j $

\item If $ \forall j: M+1, M+2,\ldots, N-1;\ x_{j}=x_{j+1}=\cdots=x_{N} $ then $\ x_j= y_{M-N+j}= \cdots =y_{M-1}=y_M $

\end{description}\begin{equation}\end{equation}

For checking previous equations (2) and (3), it is necessary to
determine the value of $N$, which depends on $k$ that is the
maximum length of possible runs of decimations. For example, if
$k=1$, then $N=3M/2$, which is the mathematical expectation of $N$
in such a case.

Note that the checking procedure of hypothesis described with the
previous Equations, applied on the intercepted sequence takes
polynomial time as it implies a simple verification of runs. The
previous patterns allow discovering promising initial states
producing sequences with a low edit distance. In fact, such a
pattern provides a good quality threshold for the method that will
be described in the following Section.

\section{General Attack}

The threshold obtained through the pattern described in the
previous Section is a fundamental ingredient of the general attack
described below. The algorithm here developed also makes use of a
new concept, the so-called {\it stop column}, which leads to a
considerable saving in the computation of the edit distance
matrices. Indeed, a {\it stop column} with respect to a threshold
$T$ may be defined as a column $j_0$ of the edit distance matrix
$W$ such that each one of their elements fulfills the Equation
(4).
\begin{equation}
\forall i \; w_{i,j_0}> T-(N-M-i), 
\end{equation}

Once a minimum edit distance threshold has been obtained, we may
use such a threshold to stop the computation of any matrix $W$ as
soon as a {\it stop column} has been detected. This is due to the
fact that the edit distance corresponding to the candidate initial
state will be worse than the threshold. In this simple way, two
new improvements on the original attack may be achieved. On the
one hand, as yet mentioned, the computation of any matrix may be
stopped as soon as a {\it stop column} is obtained. On the other
hand and thanks to the association between bits $x_{i+j}$ and
edges of the graph, we may define a new anti-pattern on the
initial states of the  target $LFSR$, the so-called
IS-anti-pattern. This  pattern is defined from the bits that produce a {\it stop column}. It allows us to discard the set
of initial states fulfilling such an IS-anti-pattern when an early
{\it stop column} has been detected. This is so because once a
{\it stop column} has been obtained, it is possible to discard
directly all the initial states whose first bits coincide with
those that produce the {\it stop column}. In order to take full
advantage of {\it stop columns} it is convenient to have some
efficient way of obtaining a good threshold, which allows us to cut many branches of the search tree.  That is exactly the
effect of the pattern described in the previous Section.

Since it is possible that the described pattern correspond only to
sequences that may not be produced by the target $LFSR$, in
practice it is convenient to restrict the pattern to the length of
the target $LFSR$. So, the pattern obtained from the first
formulas of Equations (2) and (3) limited to the length of the
target $LFSR$ is what we call IS-pattern. On the other hand,
although sequences generated through the IS-pattern have minimum
edit distance, it is possible that the corresponding obtained
decimation or insertion sequences and noise sequences are not
consistent with the description of the attacked generator. This is
the reason why the proposed algorithm includes a process of
hypothesis relaxation, which implies the successive
complementation of bits of the IS-pattern until getting a positive
result.

Finally, since the IS-pattern is determined by the runs at the
beginning of the intercepted sequence, if no long run exists at
the beginning of the sequence, initially the algorithm discards
the first bits in the intercepted sequence before a long run, and
uses those discarded bits to confirm the result of the attack. This
idea is expressed within the algorithm by a parameter $H\in
[0,L]$, chosen by the attacker depending on its computational
capacity (the greater capacity, the fewer $H$).

The full description of the proposed general edit distance attack
is as follows.
\\

{\bf Algorithm}

{\it Input:} The intercepted keystream sequence and the feedback
polynomial of the target $LFSR$ of length $L$.

{\it Output:}  The initial states of the target $LFSR$ producing
sequences with a low edit distance with the intercepted sequence,
and the corresponding decimation or insertion sequence and noise
sequence.
\begin{enumerate}

\item Verification of hypothesis on the intercepted sequence described in Equation (2) or (3).

\item While fewer than $H$ hypothesis are fulfilled, discard the first bit and consider the resulting sequence as new intercepted sequence.

\item Definition of the IS-pattern according to the first $L$ formulas in Equation (2) or (3).

\item Initialization of the threshold $T=N$.

\item  For each initial state fulfilling the IS-pattern, which has not been previously rejected:

\begin{enumerate}
\item Computation of the edit distance matrix, stopping after detecting a {\it stop column} according to threshold $T$ and Equation (4).

\item Definition of the IS-anti-pattern and rejection of all initial states fulfilling it.

\item Updating of the threshold $T$.
\end{enumerate}

\item For each initial state producing a sequence with minimum edit distance:

\begin{enumerate}
\item Computation of the shortest paths from the graph induced by the edit distance matrix.

\item Translation from each shortest path into decimation or insertion sequences and noise sequences.

\item Checking that the obtained decimation or insertion sequences, and noise sequences are consistent with the attacked generator. Otherwise, updating of the IS-pattern by complementing one of the bits in the original IS-pattern.

\end{enumerate}
\end{enumerate}

Note that if the output is not the minimum edit distance sequence,
the obtained edit distance can be used as threshold for the stop
column method in order to find such a sequence quickly.

\section{Attack on Shrinking and Alternating Step Generators}

In this Section a specific implementation of the general attack
presented in the previous Section for the cases of the $SG$ and
the $ASG$ is considered.

One of the first questions that have to be taken into account in
both cases is the limitation on the number of consecutive
deletions because the longest run of consecutive deletions in $X$
to get $Y$ is always shorter than the length $L_S$ of the selector
register $S$. This restriction implies that the equation (1)
corresponding to the computation of the edit distance matrix
should be modified in the following way:

\begin{center}
$w_{i,1}= P_i(x_{i+1},y_1),\ i=0,\ldots,L_S$

$w_{0,j}= w_{0,j-1}+P_0(x_j,y_j),\ j=2,\ldots,M$

$w_{i,1}= \infty,\ i=L_S +1,\ldots,N-M$

$w_{i,j}=\min_{k=0,\ldots, L_S -1} \{w_{i-k,j-1} +P_k(x_{i+j},
y_j)\},\ i=1,\ldots,N-M,\ j=2,\ldots,M$

$P_k(x_{i+j},y_j)=\left\{ \begin{array} {ccc}
  k &
 {\bf \rm if} & x_{i+j}=y_j\\
  k+1 & {\bf \rm if}
& x_{i+j}\neq y_j
\end{array} \right .
k=0,...,L_S -1$
\begin{equation}\end{equation}
\end{center}

Equations (2) and (3) corresponding to the definition of the
pattern in the first and the second model, respectively must be
also adapted to the $SG$ and the $ASG$, producing the Equations
(6) and (7) respectively:

\begin{description}

\item If $\forall j: 2, 3, \ldots, N-M+1; y_{1+j /  L_S}=
\cdots =y_{j-1}=y_{j}$ then $y_j= x_{L_S  (j / L_S)}=x_{L_S(j /
L_S)+1}=\cdots=x_{L_S  (j / L_S)+L_S -1}$

\item If $\forall j:N-M+2, N-M+3,\ldots, M; y_{M-N+j}=
\cdots=y_{j-1}=y_j$ then $\ y_j=x_{j}=x_{j+1}= \cdots=x_{j+L_S -1
}$

\item If $\forall j: M+1, M+2, \ldots, N-1;\ y_{M-N+j} =
\cdots =y_{M-(N-j)/ L_S}$ then $y_M=
x_j=x_{j+1}=\cdots=x_{\min(j+L_S -1, N)}$

\end{description}\begin{equation}\end{equation}

\begin{description}

\item If $\forall j: 2, 3, \ldots, N-M+1; x_{L_S  (j / L_S)}=x_{L_S(j
/  L_S)+1}=\cdots=x_{L_S  (j / L_S)+L_S -1}$ then  $ x_{L_S (j /
L_S)}= y_{1+j / L_S}= \cdots= y_{j-1}=y_{j}$

\item If $\forall j:N-M+2, N-M+3,\ldots, M; x_{j}=x_{j+1}= \cdots=x_{j+L_S
-1 } $ then $\ x_{j}=y_{M-N+j}= \cdots=y_{j-1}=y_j $

\item If $\forall j: M+1, M+2, \ldots, N-1;\ x_j=x_{j+1}=\cdots=x_{\min(j+L_S -1, N)} $ then $ x_j=y_{M-N+j} =
\cdots =y_{M-(N-j)/ L_S}
$

\end{description}\begin{equation}\end{equation}

Finally, the process of hypothesis relaxation explained in the
last Section must also be used for the cases of $SG$ and $ASG$
when the minimum obtained edit distance is greater than $N-M$
since it corresponds to the presence of noise.

The following toy example is used simply to show some of the most
remarkable aspects of the proposal.

\vspace{0,5 cm}

$Example:$

Given the situation where an attacker has intercepted the
keystream sequence $Y$:1011110 of length $M$=7 produced by a $SG$,
and knows the following parameters:

\begin{itemize}

\item $L_S=3$ and $L_A=7$

\item $P_S(x)= 1+x+x^3$ and $P_A(x)= 1+x+x^7$

\end{itemize}

If we consider $H=3$, after the verification of hypothesis from
Equation (2) on the intercepted sequence we find that none of them
is fulfilled, so we discard the first bit and check the hypothesis
on the resulting sequence. In the sequence of length 6 only two
hypotheses are fulfilled, so again we discard the first bit and
consider the resulting sequence as new intercepted sequence
$Y$:11110 of length $M$=5 because it fulfills 3 hypothesis.

If we consider $N$=10, the IS-pattern according to the first
formulas in Equation (2) corresponding to the cut sets shown in
Fig. 6 for this new sequence $X$ is given by IS-pattern: 111111x

\begin{figure}[ht]
  \begin{center}
     \includegraphics[bb=0 0 3601 2394,width=4.5in]{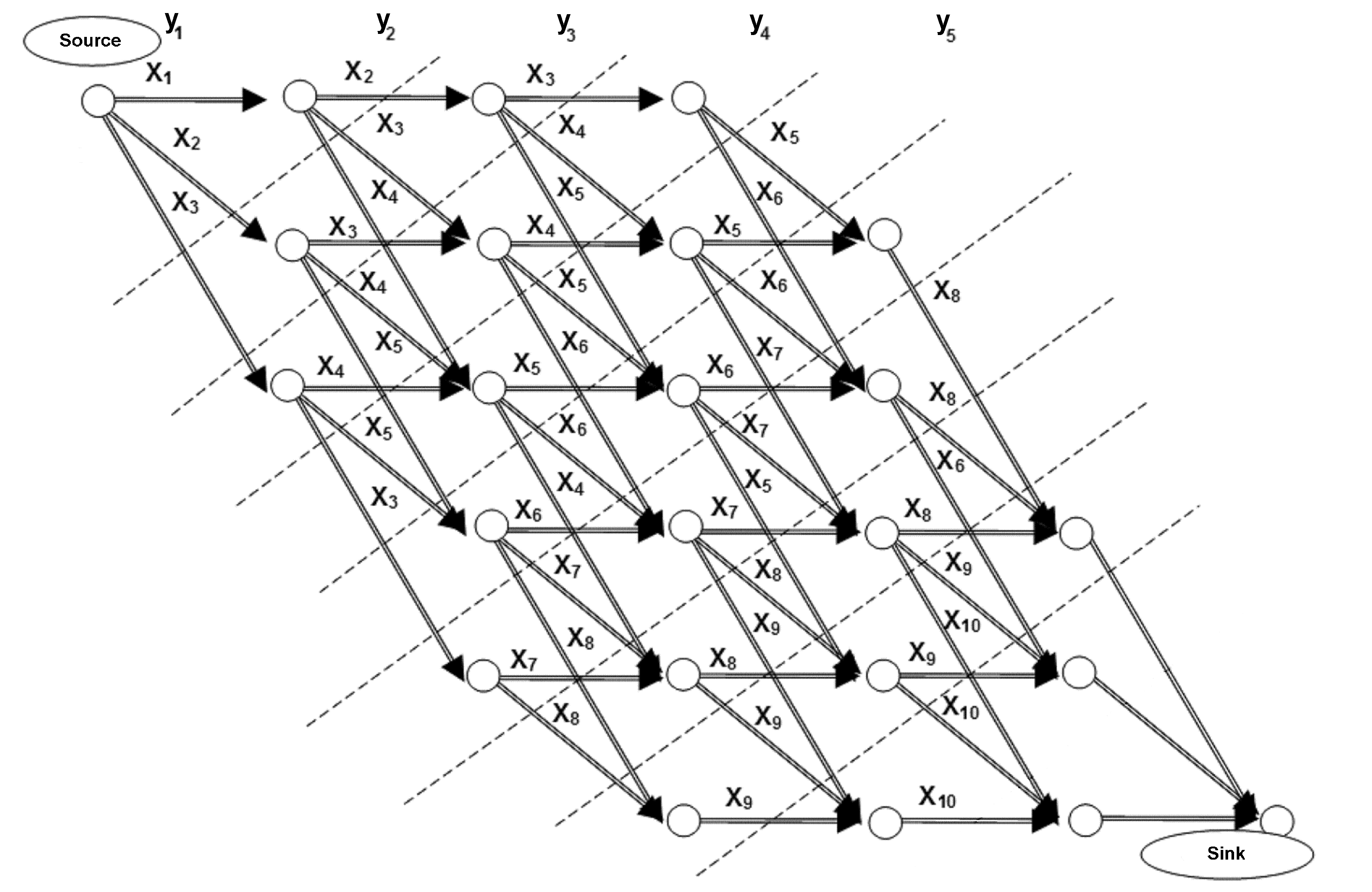}
      \end{center}
 \caption{Cut sets}
\end{figure}

Given $Y$:11110, for each initial state fulfilling the IS-pattern,
the edit distance matrix is computed:
\begin{itemize}

\item Initial state: 1111110, $X$:1111110101 and
$W= \left(
\begin{array} {ccccc}
  0 & 0 & 0 & - & - \\
  1 & 1 & 1 & 1 & - \\
  2 & 2 & 2 & 2 & - \\
  - & 3 & 3 & 4 & 4 \\
  - & 4 & 5 & 4 & 4 \\
  - & - & 5 & 6 & 6
\end{array}\right)$.

The threshold is updated by the edit distance $T=5$

\item Initial state: 1111111,$X$:1111111010 and
$W= \left(
\begin{array} {ccccc}
  0 & 0 & 0 & - & - \\
  1 & 1 & 1 & 1 & - \\
  2 & 2 & 2 & 2 & - \\
  - & 3 & 3 & 3 & 3 \\
  - & 4 & 4 & 5 & 5 \\
  - & - & 6 & 5 & 5

\end{array}\right)$.

\end{itemize}

For both initial states we get the same minimum edit distance
$N-M$=5. However, when we recover the shortest paths from the
source to the sink in the graphs induced by the edit distance
matrices, we get the following. On the one hand, in the first
case, the shortest paths correspond to non-possible decimation
sequences according to the parameters of the attacked generator.
On the other hand, the second initial state provides fifty-four
shortest paths, and only two of them correspond to possible
decimation sequences $S$: 0011101001 and $S$:1001110100 that are
consistent with $P_S$. However, if we try to confirm these
solutions with the first two discarded bits of the intercepted
sequence, we get that none of them are consistent with those bits.

Consequently, according to the algorithm, the IS-pattern has to be
updated by complementing any of the bits in the original
IS-pattern. For instance, consider the new IS-pattern: 111110x.
For this new IS-pattern again there exist two initial states
fulfilling the IS-pattern for which minimum edit distances equal
to 5 are computed. However, only the second initial state 1111101
provides a valid decimation sequence $S$: 0011101001 that is
consistent both with the parameters of the generator and with
discarded bits, so this solution is accepted as a valid solution
for the cryptanalysis. On the other hand, if we update the
IS-pattern by complementing for instance the first bit instead of
the last bit, we obtain another new IS-pattern: 011111x, and the
resulting initial state and decimation sequence are 0111111 and
$S$:0100111010, respectively.

In conclusion, the attack of this example has been successful
alter the evaluation of only 4, out of 128, promising initial
states.

\section{Simulation Results}

The next table shows some results for experimental sequential
implementations of the algorithm against shrunken sequences. The
columns denoted Seq.pat. display the number of sequences that
fulfill the IS-pattern. Cases marked with * indicate the existence
of initial states fulfilling the IS-pattern and producing
sequences $X$ that are solutions. Thres. and Dist. are the columns
where the obtained threshold and the minimum edit distance are
shown.

From these randomly generated examples, we may deduce a general
classification of inputs into several cases. The best ones
correspond to IS-patterns which directly identify solutions. On
the contrary, bad cases are those `missing the event' cases in
which the IS-pattern fails to identify any correct initial state.
Such cases are generally associated with long runs at the
beginning of the sequences $Y$. Finally, the medium cases are
those for which, despite the non existence of solutions fulfilling
the pattern, a good threshold is obtained. Such cases allow a good
percentage of saving in computing thanks to the detection of many
early {\it stop columns}.
\\
\begin{center}
\begin{tabular}{|c|c|c|c|c|c|c|} \hline
 N & M & $L_A$ & $2^{L_A}$ & Seq.pat. &
Thres. & Dist.  \\ \hline \hline
  20 & 15 & 7 & 128 &  0 &
  - & 5   \\ \hline
   30 & 20 & 9 & 512 &  1 &
   11 & 10   \\ \hline
   33 & 22 & 7 &  128 & 2* &
   12 & 12  \\ \hline
  75 & 50 & 7 &   128 & 8 &
  29 & 27  \\ \hline
   150 & 100 & 9 & 512 &  32  &
   57 & 55  \\ \hline
 300 & 200 & 11 &  2048 &  128 &
 166 & 164  \\ \hline
 300 & 200 & 13 & 8192 & 128 &
 115 & 114  \\ \hline
 450 & 300 & 13 &  8192 & 128 &
 176 & 171   \\ \hline
450 & 300 & 14 &  16384 & 1024 & 173 & 171
\\ \hline
450 & 300 & 16 & 65536 & 256  & 173 & 171
\\ \hline
750 & 500 & 14 &  16384 &  1024* & 291 & 291
\\ \hline
 \hline
\end{tabular}
\\
\end{center}

\begin{figure}[ht]
  \begin{center}
     \includegraphics[bb=0 0 280 224,width=4in]{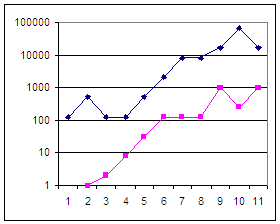}
      \end{center}
  \caption{Logarithmic plot of simulation results}
\end{figure}

Figure 7 shows a logarithmic plot with a comparison between $2^{L_A}$ and Seq.pat. obtained from the  simulation. According to those data we may estimate that the proposed algorithm produces the
solution in $O(2^{L_A /2})$ time instead of the $O(2^{L_A})$ time
corresponding to the exhaustive search that implied the original
attack \cite{P01}. These positive results are consequence of early detected stop columns. Furthermore, it is clear that the worst outputs
appear when the initial results in steps 1 to 4 are not adequate
as there are no initial states fulfilling the IS-pattern. However, we conjecture that
even in these cases that require more computation, it is
guaranteed that the solution is always obtained.  Even though it depends on the obtained pattern, from the number of branches that are cut off we may estimate that the average reduction of states to be explored is 96\%. This is reflected in a reduction of more than 25\% in the time complexity of the attack.  

Note that as aforementioned, the proposed algorithm not always
output the minimum edit distance sequence (cases that are here
denoted by an *) but however, since the hypothesis on $Y$ are
independent, the groups of bits in the IS-pattern are also
independent and consequently, the conditions might be considered
separately in such way that we might define in this way a relaxed
IS-pattern which might lead to sequences that fulfill them. In
addition, we may identify some characteristics of patterns and cases where the improvement is more dramatic. In particular, empirical results have shown that intercepted sequences
$Y$ with short runs at the beginning cause a greater improvement
in the time complexity of the attack. Thus, another way to avoid a
bad behaviour of the original algorithm is by choosing
sub-sequences from the intercepted sequence $Y$ that have no too
long runs at the beginning, and by applying the algorithm to each
one of these sub-sequences.

\section{Conclusions}

In this work a new deterministic approach to the cryptanalysis of
$LFSR$-based stream ciphers has been proposed. In particular, a
practical improvement on the edit distance attack on certain
clock-controlled $LFSR$-based generators has been proposed, which
reduces the computational complexity of the original attack
because it does not require an exhaustive search over all the
initial states of the target $LFSR$.

The main tool used for the optimization of the original attack was
the definition of graphs where optimal paths provide cryptanalytic
results and of cut sets on them that have been used to obtain a
useful threshold to cut the search tree.

An  extension of this article, which is being part of a work in
progress, takes advantage of the basic idea of using cut sets to
improve edit distance attacks against generalized clock-controlled
$LFSR$-based generators. In order to do this, the three edit
operations are considered and the resulting cut sets on the
corresponding induced graph allow the identification of promising
initial states.

\end{document}